\documentclass[11pt]{article}
\input epsf.tex
\newdimen\SaveWidth \SaveWidth=\textwidth
\newdimen\SaveHeight \SaveHeight=\textheight
\advance\SaveWidth by -\textwidth
\advance\SaveHeight by -\textheight

\divide\SaveWidth by 2
\divide\SaveHeight by 2
\advance\hoffset by \SaveWidth
\advance\voffset by \SaveHeight

\def\ie{\it i.e.}
\def\eg{\it e.g.}
\def\edm{\it edm}
\def\etc{\it etc.}
\def\starprod{\star~\rm{product}}
\def\nc{\rm noncommutative}
\def\ncg{\rm noncommutative~geometry}

\def\cpviol{CP ~\rm{violation}}
\def\cpviolng{CP ~\rm {violating}}

\def\abs#1{\left| #1\right|}

\let\badcite=\cite
\def\cite{~\badcite}

\def\slashchar#1{\setbox0=\hbox{$#1$}           
   \dimen0=\wd0                                 
   \setbox1=\hbox{/} \dimen1=\wd1               
   \ifdim\dimen0>\dimen1                        
      \rlap{\hbox to \dimen0{\hfil/\hfil}}      
      #1                                        
   \else                                        
      \rlap{\hbox to \dimen1{\hfil$#1$\hfil}}   
      /                                         
   \fi} 
    \def\slashword#1{\setbox0=\hbox{$#1$}        
  \dimen0=\wd0                                   
   \setbox1=\hbox{/} \dimen1=\wd1                
   \ifdim\dimen0>\dimen1                         
      \rlap{\hbox to \dimen0{\hfil\bf---\hfil}} %
      #1                                         %
   \else                                         
      \rlap{\hbox to \dimen1{\hfil$#1$\hfil}}    
      /                                          
    \fi}                                         %

\catcode`@=11
\newdimen\vbigd@men                             

\def\vbig#1#2{{\vbigd@men=#2\divide\vbigd@men by 2%
   \hbox{$\left#1\vbox to \vbigd@men{}\right.\n@space$}}}
\catcode`@=12

\catcode`@=11
\def\citenum#1{\csname b@#1\endcsname}
\catcode`@=12

\def\dofig#1#2{\centerline{\epsfxsize=#1\epsfbox{#2}}}
\begin{document}
\begin{titlepage}
\rightline{LBNL-50207}
\rightline{TUHEP-TH-03143}

\bigskip\bigskip

\begin{center}{\Large\bf\boldmath
Review of the Phenomenology of Noncommutative Geometry \footnotemark \\}
\end{center}
\footnotetext{ This work was supported by the Director, 
Office of Science, Office
of Basic Energy Services, of the U.S. Department of Energy under
Contract DE-AC03-76SF0098, and by the Department of Physics at Tsinghua University.
}
\bigskip
\centerline{\bf I. Hinchliffe$^{a}$ , N. Kersting$^{b}$,
		 and Y.L. Ma$^{b}$}
\centerline{{\it$^{a}$ Lawrence Berkeley National Laboratory, Berkeley, CA}}
\centerline{{\it$^{b}$ Theoretical Physics Group, Department of Physics
		       Tsinghua University, Beijing P.R.C. 100084
		       }}
\bigskip

\begin{abstract}

	We present a pedagogical review of particle physics models that are based on the noncommutativity
	of space-time, $[\widehat{x}_\mu,\widehat{x}_\nu]=i \theta_{\mu \nu}$,
 with specific attention to the phenomenology these models predict in 
	particle experiments either in existence or under development. We
	summarize results obtained for high energy scattering such as would occur for
	example in a future $e^+e^-$ linear
	collider with $\sqrt{s} = 500~GeV$,  as well as low energy experiments such
	as those pertaining to elementary electric dipole moments and other $\cpviolng$
	observables, and finally comment on the status of 
	phenomenological work in cosmology and extra dimensions.

\bigskip        

\end{abstract}

\tableofcontents

\newpage
\pagestyle{empty}

\end{titlepage}

\section{What is Noncommutative Physics?}
\label{sec:intro}
The number of papers on $\nc$ physics has been increasing rapidly in the 
past 15 years. This is undoubtably due to the recent ground-breaking work in string theory
 demonstrating how $\ncg$ can arise in the context of $D$-branes \cite{Witten:1986cc,
Seiberg:1999vs}; as the
string theory community refines the fundamentals of $\ncg$, model-building efforts 
directed at experimental detection of $\nc$ effects grow more numerous. This
 review attempts to introduce to the non-specialist the extensive work on $\nc$ 
phenomenology in the literature and provide a summary of the work in progress and
challenges ahead. This section will define the key ideas 
necessary to understand 
only the phenomenological work on $\ncg$. For the
reader interested in the more formal side of the theory there is a large body of papers
available at an accessible level (see, for example, \cite{madore99,
Connes:2000ti, Kastler:2000ff, Carlen:2001wj, Varilly:1997qg,Girotti:2003at}).

To begin, we must define what we mean by ``$\nc$ space-time'':
Noncommutative space-time\cite{Connes:2000ti} is a deformation of ordinary space-time in which the
 space-time coordinates $x_\mu$, representable by Hermitian operators $\widehat{x}_\mu$, do not
commute:
\begin{equation}
\label{nceqn}
[\widehat{x}_\mu,\widehat{x}_\nu]=i \theta_{\mu \nu}
\end{equation}
Here $\theta_{\mu \nu}$ is the deformation parameter: ordinary space-time is obtained in the
 $\theta_{\mu \nu} \to 0$ limit.  By convention
it is a real tensor\cite{Cai:2000py} antisymmetric under $\mu \leftrightarrow \nu$. 
Note that $\theta_{\mu \nu}$ has dimensions 
of length-squared; the physical interpretation of this is that  $\theta_{\mu \nu}$
is the smallest patch of area in the ${\mu \nu}$-plane one may deem ``observable,'' analogous
 to the role $\hbar$ plays in 
$[\widehat{x_{i}},\widehat{p_{j}}]=i \hbar \delta_{ij}$,
 defining the corresponding smallest
patch of observable phase space in quantum mechanics. Noncommutative space-time as defined
above is therefore one parameterization of the supposed limit to the resolution with which
one may probe space-time itself. This type of space-time is often called ``fuzzy''\cite{Ydri:2001pv},
as there is no definite meaning to a ``point''; in $\nc$ quantum field theory (see Section 
\ref{subsec:qft} below)
interactions between fields at distinct points are ``smeared'' by a  $\theta$-dependent
 weighting function. Noncommutative phenomenology in this review will involve
 calculating observables to $\cal{O}(\theta)$ unless otherwise noted.

Noncommuting coordinates are not new in physics: a nonrelativistic charged particle in a strong
magnetic field, for example, moves in a noncommuting space. Taking the particle's motion to be in the $x-y$ 
plane in the presence of a $B$-field pointing in the $z$-direction, the Lagrangian is
\begin{equation}
L={1\over 2} m v^2 + {e\over c}v\cdot A - V(x,y)
\end{equation}
where $m$ and $e$  are the paricle's mass and charge, respectively, 
$A_\mu=(0,0,0,Bx)$ and $V$ is some conservative potential. Taking the strong field limit $B \to \infty$, 
this Lagrangian becomes
\begin{equation}
L= \frac{e B}{c}x \dot{y} - V(x,y)
\end{equation}  which is of the form $p \dot{q} - h(p,q)$ if we identify 
$(\frac{eBx}{c},y)$ as canonical coordinates,
$\ie$ $\{x,y\} = \frac{c}{eB}$
 \cite{Jackiw:2001dj,Dvoeglazov:2002sc,Magro:2003bs}.
 This behaviour is closely
related to the origin of $\ncg$ in string theory \cite{Witten:1986cc}). More to the point,
$\ncg$ is expected on quite general
grounds in any theory that seeks to incorporate gravity into 
a quantum field theory. Semi-classically, to localize
a particle to within a Planck length $\lambda_P$ in any given plane, an energy equal to the 
Planck mass must be available to the particle; this creates a black hole that
 swallows the particle. To avoid this, one must insist 
 $\sum_{i<j}{\Delta x_i \Delta x_j} \geq {\lambda_P}^2$; alternatively,
one is led to think of space as a
noncommutative algebra upon trying to quantize the 
Einstein theory\cite{Dubois-Violette:1998vq,Doplicher:1995tu}. Noncommuting space-time coordinates
are unobservable in this case as ${\lambda_P}^{-1} \approx 10^{19}~GeV$, but
 they may have visible effects if $\theta$ is large 
or if ${\lambda_P}^{-1} \approx 1~TeV$ as is possible in theories with 
extra dimensions (see \ref{subsec:cosmosdim} below).

 In this review we consider observables in two cases: high energy collider and
low energy precision experiments. The former includes $e^+e^-$ linear and hadronic colliders, whereas the
 latter includes atomic experiments, anomalous magnetic moment and electric dipole moment 
measurements, and other CP-violation tests.
After introducing the basics of $\nc$ quantum field theory in Section \ref{sec:ncsm},    
 we present in Section \ref{sec:phen}  the status of the main areas of this 
phenomenological work in $\ncg$.
 We will also mention cosmological observables and the possibility of noncommuting extra dimensions,
and in Section \ref{sec:chall} we shall indicate the directions in which work 
in phenomenological $\ncg$ needs more attention.

\section{Status of a Noncommutative Standard Model}
\label{sec:ncsm}

At present, there is no consensus in the literature as to the precise form of the
Noncommutative Standard Model~(NCSM) due to some theoretical difficulties which
we will discuss below.
 However there is very promising work underway\cite{Calmet:2001na}
which seems to explicitly show that the NCSM is realizable, albeit by some rather nontrivial 
adjustments to the naive model (see \ref{subsec:gauge} below). Most of the work in $\nc$
phenomenology assumes that these difficulties will be eventually 
addressed, perhaps motivated by the confidence that the low
energy theory (the Standard Model) and the high energy theory (String Theory) are both believable 
candidates for descriptions of Nature, and that a NCSM consisting of aspects of both 
should be a theory which is valid 
for intermediate energies\cite{zumino}.
 We now present some of the essential features of a $\nc$ quantum field theory.

\subsection{Quantum Field Theory Details}
\label{subsec:qft}

\subsubsection{Fundamentals}

We define $\theta$ to be the average magnitude of an element of $\theta_{\mu \nu}$; physically 
  ${1\over\sqrt{\theta}}$ corresponds to the energy threshold beyond which a particle moves and interacts in
a distorted space-time. We expect the effects of the deformation 
 $\theta_{\mu \nu}$ to be somehow manifest at energies $E$ well below   ${1\over\sqrt{\theta}}$, where
an effective operator expansion in the small parameter $E \sqrt{\theta}$ is feasible.
 Essentially all the studies we
present subsequently approach the theory in this manner.  

Working in $\nc$ field theory is equivalent to working with 
ordinary (commutative) theory and
replacing the usual
product by the $\starprod$ defined as follows:
\begin{equation}
\label{star}
(f \star g)(x) \equiv e^{i \theta_{\mu \nu} \partial_{\mu}^{y} \partial_{\nu}^{z}}
		f(y) g(z) \mid_{y=z=x}
\end{equation}
With this definition (\ref{nceqn}) holds in function space equipped with a $\starprod$:
\begin{equation}
\label{nceqn2}
[x_{\mu},x_{\nu}]_{\star}=i \theta_{\mu \nu}
\end{equation}
Note that when $f(x)$ and $g(x)$ are fields the $\starprod$ in Equation \ref{star}
 can be  written in momentum space as
\begin{equation}
\label{starmoment}
(f \star g)(x) \to f(p)g(q) e^{ip^\mu \theta_{\mu\nu}q^\nu} \equiv f(p)g(q) e^{ip \times q}
\end{equation}
where we have defined $p \times q \equiv p^\mu \theta_{\mu\nu}q^\nu$. 
This $\starprod$ intuitively replaces the point-by-point multiplication of two fields
by a type of ``smeared'' product. More detailed analysis of 1- and 2- point functions\cite{Chaichian:1998kp}
bears out this intuition:
spacetime is only well defined down to distances of order $\sqrt{\theta}$ so
functions of spacetime must be appropriately averaged over the appropriate neighborhood.
 More precisely, in each $(i,j)$ plane, we must replace 
\begin{equation}
\label{avg}
\phi(x_i,x_j) \rightarrow \int dx_i'dx_j' \phi(x_i',x_j') e^{-\frac{(x_i-x_i')^2+(x_j-x_j')^2}{\theta_{ij}}}(\pi \theta_{ij})^{-1} 
\end{equation} 

Making predictions with a $\nc$ model therefore appears straightforward: just replace each ordinary product
with a $\starprod$. However, a number of difficulties arise with this replacement which
 make the task of formulating a Noncommutative Standard Model (NCSM) not entirely trivial. In this 
section we review these difficulties and the solutions found thus far.

As noted above, the method of computing $\nc$ field theory amplitudes is effected by replacing
the ordinary function product with the $\starprod$ in the 
Lagrangian. The theory is otherwise identical to the
commuting one (i.e. the Feynman path integral formulation
provides the usual setting for doing quantum field theory).
In practice it is simpler to work in momentum space, where the only modification to the Feynman rules 
is to replace each vertex factor $\lambda$ with a momentum-dependent factor 
$\lambda V(p_1,p_2,...p_n)$ with
\begin{equation}
V(p_1,p_2,...p_n)=e^{-\frac{i}{2} \Sigma_{i<j}p_i \times p_j}
\end{equation}
Here $p_i$ is the momentum flowing into the vertex from the $i$-th field. In this formulation where 
noncommutivity introduces pairwise contractions of momenta  in Feynman diagram amplitudes
($\ie$ factors of $e^{i p\times q}$, see  (\ref{starmoment}))
 it is clear that some of these contractions are trivial, $\eg$
 $e^{i q \times q} = 1$ since $\theta_{\mu\nu}$ is antisymmetric. In particular when the momenta 
associated with the internal lines of a given Feynman diagram are contracted and 
only the contraction of the external momenta remain, we refer to this as a $SP$(``simply-phased'') 
diagram, since it is equal to the ordinary diagram apart from an overall $\theta$-dependent
phase factor; otherwise, the diagram is NSP(``non-simply-phased'')\footnotemark.
In this characterization it is the
NSP Feynman diagram which typically modifies the ultraviolet behaviour of the theory in a nontrivial way 
(see the example in Section \ref{subsub:iruv} below). 
\footnotetext{in the literature these are often called ``planar'' and 
``nonplanar'' diagrams \cite{Filk:1996dm,Minwalla:1999px} }

  There has already been extensive work on
scalar field theory\cite{VanRaamsdonk:2000rr,Grosse:1996mz,Aref'eva:1999sn}, 
NCQED (the \linebreak $\nc$  analog of
 QED)\cite{Hayakawa:1999zf,Nguyen:2003ry},
as well as noncommutative Yang-Mills\cite{Krajewski:1999ja,Armoni:2000xr}; perturbation theory in $\theta$ 
is applicable and the theories are renormalizable\cite{Bonora:2000ga,Brouder:2000er}. 
For example
a Yukawa theory with a scalar $\phi$, Dirac fermion $\psi$, has
the action
\begin{equation}
\label{yukawa}
S=\int{ d^4 x \left( \overline{\psi} i \slashchar{\partial} \star \psi
+ \partial_\mu \phi \star \partial^\mu \phi + \lambda \overline{\psi} \star \psi \star \phi \right)}
\end{equation}
which to ${\cal O}(\theta)$ is
\begin{equation}
\label{yukawa2}
S=\int{ d^4 x \left( \overline{\psi} i \slashchar{\partial} \psi
+ (\partial_\mu \phi)^2  + \lambda \left(  \overline{\psi} \psi \phi +
\partial_\mu \overline{\psi} \theta^{\mu\nu} \partial_\nu \psi \phi +
\partial_\mu \overline{\psi} \theta^{\mu\nu} \partial_\nu \phi \psi +
 \overline{\psi} \partial_\mu \psi \theta^{\mu\nu} \partial_\nu\phi    
\right)
\right)}
\end{equation}

(Here we have used the fact that $\int{ dx \xi \star \xi} = \int{ dx \xi \xi}$,
which follows straightforwardly from (\ref{star})).
Likewise, the usual $\phi^4$ theory has an action
\begin{equation}
\label{phi4}
S=\int{ d^4 x \left( (\partial \phi)^2 + m^2\phi^2 
+ \frac{\lambda}{4!} \phi \star \phi \star \phi \star \phi \right)}
\end{equation}
which to ${\cal O}(\theta)$ is
\begin{equation}
\label{phi4_2}
S=\int{ d^4 x \left( (\partial \phi)^2 + m^2\phi^2 
+ \frac{\lambda}{4!}\left( \phi^4 + 
6(\partial_\mu \phi ~ \theta^{\mu\nu} \partial_\nu \phi)  \phi^2 \right) \right)}
\end{equation}
We see that
the interaction terms above contain derivatives coupled to powers of 
$\theta_{\mu \nu}$.
 It is these terms which will lead to mixing between 
infrared and ultraviolet limits. 

\subsubsection{IR/UV mixing}
\label{subsub:iruv}

Calculating the one-loop quadratic effective action from  (\ref{phi4}),
one needs to consider the contribution from both the $\theta$-dependent and
 $\theta$-independent interactions (see Equation \ref{phi4_2})). Equivalently, one
computes the $SP$ and $NSP$ graphs as done in \cite{Minwalla:1999px}, with the result
\begin{equation}
\begin{array}{l}
{\Gamma_1}_{SP} = \frac{\lambda}{3 (2\pi)^4}\int {\frac{d^4 k}{k^2 + m^2}} \\
{\Gamma_1}_{NSP} = \frac{\lambda}{6 (2\pi)^4}\int {\frac{d^4 k e^{i k \times p}}{k^2 + m^2}} \\
\end{array}
\end{equation}
where $p$ is the external momentum. The $SP$ integral is quadratically 
divergent, but the oscillatory factor in the $NSP$ diagram renders it ultraviolet finite. Introducing
the Schwinger\cite{Schwinger:1949ra} parameter $\alpha$ by $\frac{1}{k^2+m^2} = 
{\int_0}^\infty{d\alpha ~e^{-\alpha(k^2+m^2)}}$ and a Pauli-Vilars regulator 
$e^{-\frac{1}{M^2 \alpha}}$ now gives 
\begin{equation}
\label{gammaloop}
\begin{array}{l}
{\Gamma_1}_{SP} = 
\frac{\lambda}{48 \pi^2}\int {\frac{d\alpha }{\alpha^2}e^{-\alpha m^2 -\frac{1}{M^2 \alpha}}} 
= \frac{\lambda}{48 \pi^2} (M^2 - m^2 ln(\frac{M^2}{m^2}) + {\cal O}(1)) \\
{\Gamma_1}_{NSP} =
 \frac{\lambda}{96 \pi^2}\int {\frac{d\alpha }{\alpha^2}e^{-\alpha m^2 -\frac{p \circ p + 
1/ M^2 }{\alpha}}} 
= \frac{\lambda}{96 \pi^2} ({M_{eff}}^2 - m^2 ln(\frac{{M_{eff}}^2}{m^2}) + {\cal O}(1))
\\
\end{array}
\end{equation}
where $p \circ p \equiv \abs{p^\mu \theta_{\mu\nu} \theta^{\nu\rho} p_\rho}$ and 
${M_{eff}}^2 \equiv \frac{1}{1/M^2 + p\circ p}$, from which it is clear that in the
ultraviolet limit $M \to \infty$ the nonplanar contribution is finite. The total $1PI$ 
quadratic effective action to ${\cal O}(\lambda)$ is therefore
\begin{equation}
\label{phi4IR}
S_{1PI} = \int{ d^4 p \frac{1}{2} \phi(p) \phi(-p) \left( 
p^2 + \tilde{m}^2 + \lambda \frac{{M_{eff}}^2}{96 \pi^2}
 -\frac{\lambda M^2}{96 \pi^2} 
 ln \left(\frac{{M_{eff}}^2}{M^2}\right) \right)}
\end{equation}
where $\tilde{m}^2 = m^2 + \frac{\lambda M^2}{48 \pi^2} - \frac{\lambda m^2}{48 \pi^2}ln(\frac{M^2}{m^2})$
 is the renormalized mass. The surprising feature of 
this result is that the limits $\theta\to 0$ or $p\to 0$ (an IR limit) 
and $M \to \infty$ (a UV limit) don't commute.
Ordinarily in field theory one takes the cutoff to infinity
and performs a subtraction, but here it appears that doing so leaves IR singularities (as $p \to 0$).
The higher the cutoff, the more sensitive the amplitude becomes to the lowest energies available.
This is not in accord with a Wilsonian intuition of renormalization, in which
  low energy effective theory decouples from high energy dynamics. Hence this phenomenon is called
the ``IR/UV mixing problem''. It persists in the $\nc$ version of QED, 
NCQED. 

 It may well be that,  as pointed out in \cite{Calmet:2001na,VANRAAMSDONK:2001JD}, IR/UV may have a novel
 interpretation which is not disastrous for the theory (see for example \cite{discrete}),
 or IR/UV may simply not be
 an {\it observable} problem
 at all. It has been shown that supersymmetry may alleviate the problem:
the authors of \cite{Matusis:2000jf} find that dangerous IR/UV contributions to the photon
 propagator are proportional to the number of bosons minus the number of fermions in the
theory, $(n_F-n_B)$, which is of course zero in unbroken supersymmetry.  
 Reference \cite{Amelino-Camelia:2002au} more completely discusses the theory in broken supersymmetry.

\subsubsection{Lorentz Violation}
\label{subsubsec:lorentz}
Because $\theta_{\mu \nu}$ carries Lorentz indices, it does not violate Lorentz symmetries 
under transformations of the observer's frame, but since it is a {\it constant} object, it violates 
``particle Lorentz invariance'' \cite{Carroll:2001ws}, which means that in any given frame
it singles out a particular direction $\theta^i \approx \epsilon_{ijk} \theta^{jk}$ in space.
Many researchers set $\theta_{0i}=0$ to avoid problems with unitarity and
causality\cite{Chaichian:2000ia}, so we do not include the time components in the definition of this ``direction''.
However it is not necessary to use this constraint for the purposes of low-energy phenomenology.
As we will see below in \ref{sec:phen}, this orientation of $\theta_{\mu \nu}$ provides
 a strong constraint on the magnitude of $\theta_{\mu \nu}$.

However if $\theta$ varies over intervals of space-time much smaller than those in
which experiments operate, then there is no well-defined meaning 
to $\theta^i$ in terrestrial experiments and the aforementioned constraints 
effectively drop away.
Nevertheless, a nonzero $\theta$ will affect the dispersion relation 
$E^2 = p^2 + m^2$ of particles 
travelling through space-time. Among some of the work in Lorentz violation 
\cite{Carroll:2001ws,Konopka:2002tt,Colladay:2002bi,Jacobson:2001tu}
 are predictions of a
number of exotic phenomena, such as the 
decay of high-energy photons and charged particles
 producing Cerenkov radiation in vacuum.

\subsection{Gauge Interations and Particle Spectra}
\label{subsec:gauge}
Similar to the non-gauge theories above in (\ref{yukawa},\ref{phi4}), a noncommutative
gauge theory follows from the ordinary one by inserting the $\starprod$ everywhere. For
NCQED, for example, we have the action
\begin{equation}
\label{ncqed}
S=\int{ d^4 x \left(  -\frac{1}{4e^2}F^{\mu \nu} F_{\mu \nu} +  \overline{\psi}  
  i\slashchar{\partial}\psi -  e \overline{\psi} \star \slashchar{A} \star \psi
 - m \overline{\psi} \psi\right)}
\end{equation}
where
\begin{equation}
\label{Fterm}
F_{\mu \nu} \equiv \partial_\mu A_\nu - \partial_\nu A_\mu - i [A_\mu,A_\nu]_\star
\end{equation}
Note the extra term in the field strength which is absent in ordinary QED; this
nonlinearity gives NCQED  a NonAbelian-like structure. There will be, for example,
3- and 4-point photon self-couplings at tree level (see Figure \ref{feyn}).  However, if one
assumes that the gauge transformation of the connection $A_\mu$ is the naive one, $\ie$
\begin{equation}
\label{transf}
A_\mu \to U \star A_\mu U^{-1} + i U \star \partial_\mu U^{-1}
\end{equation}
where $U \equiv (e^{i \lambda^a T_a})_{\star}$ for gauge parameters $\lambda^a$ and 
group generators $T^a$ , then the infinitesimal transformation of $A_\mu$ is \cite{Terashima:2000xq}
\begin{equation}
\begin{array}{ll}
\delta_{\lambda}A_\mu & = \partial_\mu \lambda + i \lambda \star A_\mu - i A_\mu \star \lambda \\
& = T^a(\partial_{\mu} \lambda_a)+ \frac{i}{2} [T^a, T^b](\lambda_a \star {A_b}_\mu + {A_b}_\mu \star\lambda_a)
+ \frac{i}{2} \{T^a, T^b\}(\lambda_a \star {A_b}_\mu - {A_b}_\mu \star\lambda_a) \\
\end{array} 
\end{equation}
This must be closed under the gauge group, so in particular $\{T^a, T^b\}$ must be a generator as
well, which is true only for $T \subset U(N)$. Therefore it appears that only $U(N)$ gauge groups
are compatible with $\ncg$. Moreover, a similar calculation shows that any matter content of the
theory that transforms in the naive way  must be in either the fundamental, anti-fundamental, 
or singlet representation of the gauge group. Other workers subsequently found that
 products of gauge
groups were likewise constrained \cite{Chaichian:2001py}. The result of recent work, however, 
claims that this is not a problem if one assumes a more complicated transformation\cite{Calmet:2001na, Jurco:2001rq}
 based on 
the Seiberg-Witten map\cite{Witten:1986cc} to replace the naive one in (\ref{transf}).

\subsection{A Working Model}
\label{sub:work}
\begin{figure}[t]
\dofig{6in}{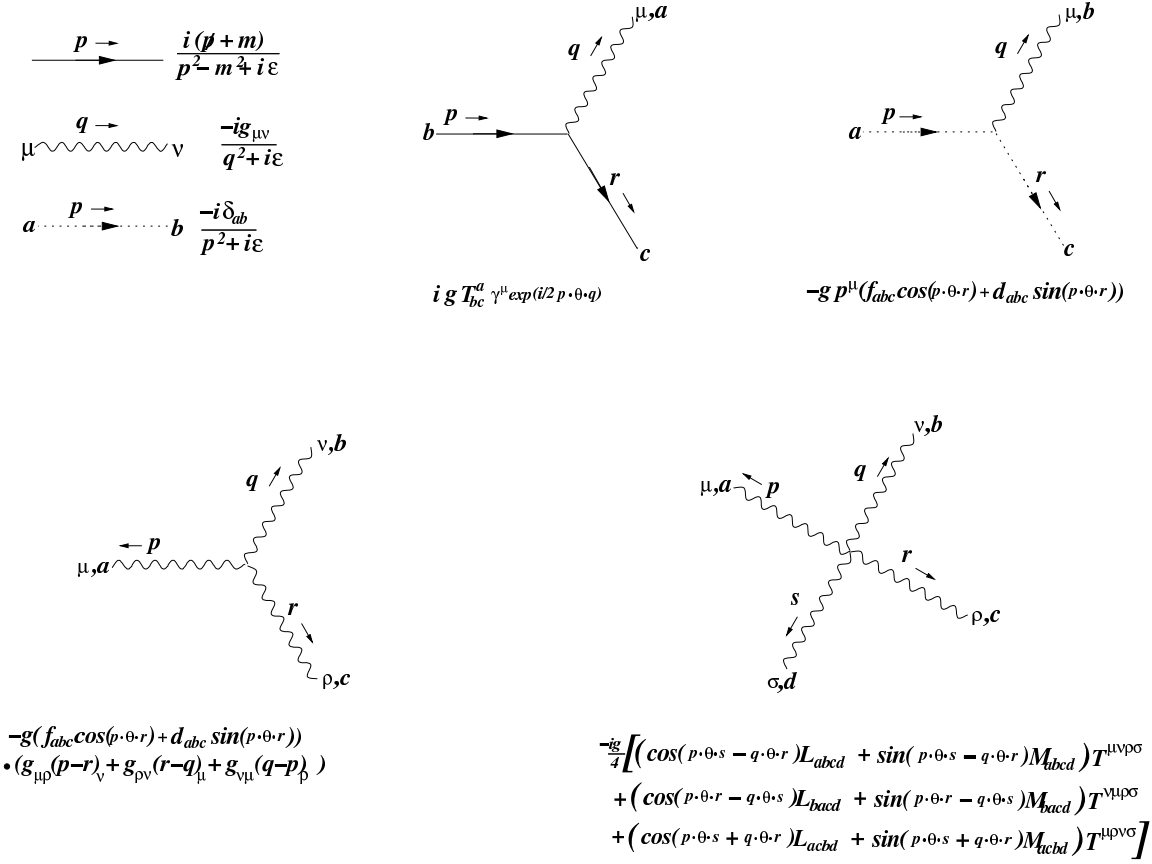}
\caption{Feynman rules for fermions (solid lines), gauge particles (wavy lines),
and ghosts (dotted lines). 
Notation: ~~~
$p,q,r,s$ Momenta ~~~~~  
$\mu,\nu,\rho,\sigma$ Lorentz indices ~~~~~
$a,b,c,d$  gauge indices ~~~~~ $T^a_{bc}$ gauge generator ~~~~~
$f_{abc}$ structure constants for $SU(N)$: $[T_a,T_b] = f_{abc}T^c$ ~~~~~~
$d_{abc}$ structure constants for $SU(N)$: $\{T_a,T_b\} = d_{abc}T^c + {1 \over N}\delta_{ab}$
~~~~~ $L_{abcd} \equiv d_{abe} d_{cde} + d_{ade} d_{cbe} - f_{abe} f_{cde} - f_{ade} f_{cbe} $ ~~~~~
 $M_{abcd} \equiv d_{abe} f_{cde} - d_{ade} f_{cbe} + f_{abe} d_{cde} - f_{ade} d_{bce}$ ~~~~~
$T_{\mu\nu\rho\sigma} \equiv g_{\mu\nu}g_{\rho\sigma} +g_{\mu\sigma}g_{\nu\rho}- 2 g_{\mu\rho}g_{\nu\sigma}$ 
~~~~ For QED/Weak vertices, index $0$ corresponds to a photon: $d_{0,i,j} = \delta_{ij}$, 
 $d_{0,0,i} = 0$, and $d_{0,0,0} = 1$, $f_{0,a,b}$ = 0.
\label{feyn} }
\end{figure}

Following the above discussion a viable NCSM is possible, though some points need further clarification. The Feynman rules 
are shown in Figure \ref{feyn}. 
The essential difference from the usual SM Feynman rules
are  the momentum-dependent oscillatory factors present at each vertex.
Otherwise, there is no change from the SM (for a detailed presentation of the
 model and related work, please see \cite{Calmet:2001na,
Chaichian:2001py,Madore:2000en,Hewett:2001im,He:2002mz}. 
For the rest of this review we assume that
it is permissible to compute in the NCSM operating under the assumption that any subsequent 
refinements of the basic theory won't drastically affect the ${\cal O}(\theta)$ expansion of
the above Feynman rules and the phenomenological
work presented below. 

\newpage

\begin{figure}[t]
\dofig{3.50in}{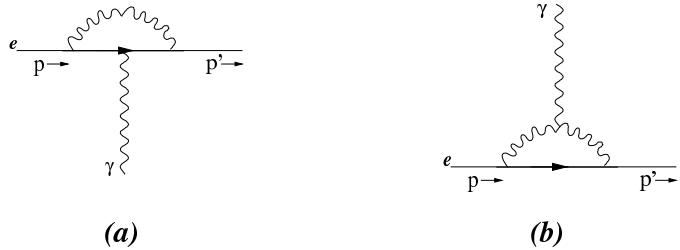}
\caption{The one-loop contributions to the electron vertex function $\Gamma_\mu$ in NCQED. Each $\gamma ee$
vertex contributes a momentum-dependent phase factor which affects the loop integral.
Whereas (a) is present in ordinary QED, (b) is a novel contribution due to $\ncg$. 
\label{edmfig} }
\end{figure}

As an example of the application of this working model, consider the diagrams in Figure \ref{edmfig} which 
contribute to the one-loop electron vertex function ${\Gamma^{(1)}}_{\mu}$ in NCQED. 
\newpage
These arise from

\begin{equation}
\label{edmloop}
\begin{array}{ll}
{\Gamma^{(1)}}_{\mu} = & {\Gamma^{(1a)}}_{\mu} + {\Gamma^{(1b)}}_{\mu} \\
{\Gamma^{(1a)}}_{\mu} = &
i (-ie)^2 e^{\frac{i}{2} p\times p'}
\int{\frac{d^4 k}{(2 \pi)^4} e^{-i k \times q}
\frac{\gamma_{\sigma}}{k^2 - \mu^2 + i \epsilon}
 \frac{\slashchar{p}' - \slashchar{k} + m}{(p'-k)^2 -m^2 + i\epsilon}
\gamma_{\mu} 
 \frac{\slashchar{p} - \slashchar{k} + m}{(p-k)^2 -m^2 + i\epsilon} \gamma^\sigma} \\
{\Gamma^{(1b)}}_{\mu} = &
- i e^2 e^{\frac{i}{2} p\times p'}\int \frac{d^4 k}{(2 \pi)^4}
\frac{1-e^{iq\times k + ip' \times p}}{(k^2-m^2)((p'-k)^2-\mu^2)((p-k)^2-\mu^2)} \\
& \cdot \left(\gamma_{\nu}(\slashchar{k} + m)\gamma_\rho [g^{\mu\nu}(2p'-p-k)^\rho +
g^{\nu\rho}(2k-p'-p)^\mu + g^{\rho\mu}(2p-p'-k)^\nu] \right)
       \\
\end{array}
\end{equation}
Here ${\Gamma^{(1a)}}_{\mu}$  is the contribution from the diagram in Figure \ref{edmfig}(a),
similar to the one in ordinary QED, while  ${\Gamma^{(1b)}}_{\mu}$ is from the
diagram  in Figure \ref{edmfig}(b), a new contribution with no counterpart in QED (note the
tri-photon coupling), $p,p'$ are external electron momenta, $k$ is the photon loop momentum,
$q=(p'-p)$ and $\mu$ is a 
small nonzero mass for the photon to regulate IR divergences \cite{Riad:2000vy}.
When we introduce the Schwinger parameters and insert a UV regulator as done above Equation \ref{gammaloop}
 (for more details, see the computation in \cite{Riad:2000vy}),
\begin{equation}
\label{edmloop2}
\begin{array}{ll}
{\Gamma^{(1a)}}_{\mu} = & -\frac{\alpha}{\pi}e^{\frac{i}{2} p\times p'}
  \int^1_0 d\alpha_1 d\alpha_2  d\alpha_3 \delta(1-\alpha_1-\alpha_2-\alpha_3)
  e^{-i(\alpha_2+\alpha_3)q\times p} \\ 
 & \cdot \left(\frac{2 A_\mu K_1 (2 \sqrt{X})}{ \sqrt{X} M^2_{eff}}
     + 2 B_\mu K_0 (2 \sqrt{X})   +2 \sqrt{X}  M^2_{eff} C_\mu K_1 (2 \sqrt{X}) \right) \\
{\Gamma^{(1b)}}_{\mu} = &  -\frac{\alpha}{\pi}e^{\frac{i}{2} p\times p'}
  \int^1_0 d\alpha_1 d\alpha_2  d\alpha_3 \delta(1-\alpha_1-\alpha_2-\alpha_3) \\
  & \cdot [  
 e^{-i(\alpha_2+\alpha_3)q\times p}  \left(\frac{2 \tilde{A}_\mu K_1 (2 \sqrt{Y})}{ \sqrt{Y} M^2_{eff}}
     + 2 \tilde{B}_\mu K_0 (2 \sqrt{Y})   + \sqrt{Y}  M^2_{eff} \tilde{C}_\mu K_1 (2 \sqrt{Y}) \right) \\
& + \left(\frac{2 \overline{A}_\mu K_1 (2 \sqrt{Z})}{ \sqrt{Z} M^2}
     + 2 \overline{B}_\mu K_0 (2 \sqrt{Z})
         \right)] \\ 
\end{array}
\end{equation}
where
\begin{equation}
\begin{array}{l}
 A_\mu =  \gamma_\mu \left(p' \cdot p - \frac{1}{2}(\alpha_2 +\alpha_3)(p' + p)^2
  + \frac{1}{2} m^2 (\alpha_2 +\alpha_3)^2 -   \frac{1}{2}\alpha_2 \alpha_3 q^2 \right)
  +  \frac{m}{2} \alpha_1 (\alpha_2 +\alpha_3)(p' + p)_\mu \\
 \tilde{A}_\mu =  \frac{1}{2} \left((\alpha_2 +\alpha_3)(p' + p)^2 - 3 m^2 - m^2 (\alpha_2 +\alpha_3)^2
         +  \alpha_2 \alpha_3 q^2  \right) + \frac{m \alpha_1}{2} (\alpha_2 +\alpha_3)(p' + p)_\mu \\
 \overline{A}_\mu =  \frac{1}{2}  \gamma_\mu \left( 
                (\alpha_2 +\alpha_3)(p' + p)^2 - 3 m^2 - m^2 (\alpha_2 +\alpha_3)^2 \right)
               +  \frac{m}{2} \alpha_1 (\alpha_2 +\alpha_3)(p' + p)_\mu \\ 
i B_\mu =  \frac{\gamma_\mu}{2 i} + \frac{\gamma_\mu q\times p}{2} (2 - \alpha_2 -\alpha_3)
   +  \frac{m}{2}(1 + \alpha_2 + \alpha_3)q^\nu \theta_{\nu \mu} 
   +  \frac{1}{4}(1 + \alpha_1)(p' + p)_\mu q \times \gamma \\
 i \tilde{B}_\mu =  \frac{3i}{2}\gamma_\mu +  
 \frac{1}{2}(mq^\nu \theta_{\nu \mu} + \gamma_\mu q \times p) (2 - \alpha_2 -\alpha_3)
 - \frac{q \times \gamma}{4}(p' + p)_\mu (1 + \alpha_2 + \alpha_3) \\
 \overline{B}_\mu = \frac{3}{2} \gamma_\mu \\
 C_\mu =  -\frac{\gamma_\mu q \circ q}{8} + \frac{q^\nu \theta_{\nu \mu} q \times \gamma}{4}\\
 \tilde{C}_\mu = \frac{q \times \gamma q^\nu \theta_{\nu \mu} }{4} + \frac{\gamma_\mu q \circ q}{8} \\
X = \frac{\alpha_1 \mu^2 + (\alpha_2 +\alpha_3)^2 m^2 - \alpha_2 \alpha_3 q^2}{M^2_{eff}} \\
Y =  \frac{ (\alpha_1 - \alpha_2 - \alpha_3)m^2 + (\alpha_2 +\alpha_3) \mu^2
          + (\alpha_2 +\alpha_3)^2 m^2 - \alpha_2 \alpha_3 q^2}{M^2_{eff}} \\
Z = \frac{ (\alpha_1 - \alpha_2 - \alpha_3)m^2 + (\alpha_2 +\alpha_3) \mu^2
          + (\alpha_2 +\alpha_3)^2 m^2 - \alpha_2 \alpha_3 q^2}{M^2} \\
\end{array}
\end{equation}
and $M^2_{eff} \equiv 1/(M^{-2} -q \circ q /4)$ and $K_0,K_1$ are modified Bessel functions of the first and 
second kind, respectively. As $M^2_{eff} \to \infty$, all of $X,Y,Z$ tend to zero, and the
 Bessel functions approach the asymptotic forms
\begin{equation}
\begin{array}{l}
K_{0}(\xi) \to - ln(\xi) \\
K_{1}(\xi) \to  1/\xi \\
\end{array}
\end{equation} 
Note in Equation \ref{edmloop2} above this implies that all the terms containing $K_1$ are finite. 
However, 
those terms containing $K_0$ (all proportional to $B_\mu$, $\tilde{B}_\mu$, 
or $\overline{B}_\mu$) logarithmically diverge; we can absorb these divergences into counterterms
proportional to each $B_\mu$, $\tilde{B}_\mu$, 
or $\overline{B}$ term, just as in the usual QED. The renormalized vertex function is
then found to contain well-behaved finite terms plus the following:
\begin{equation}
\label{iruvterms}
\begin{array}{ll}
{\Gamma^{(1)}}_{\mu} \supset &  -\frac{\alpha e^{\frac{i}{2} p \times p'}}{\pi}
   \int^1_0  d\alpha_1 d\alpha_2  d\alpha_3 \delta(1-\alpha_1-\alpha_2-\alpha_3) \\
&  \cdot M^2_{eff} \left(C_\mu e^{-i (\alpha_2 + \alpha_3) q \times p}
    - \tilde{C}_\mu  e^{i (\alpha_2 + \alpha_3) q \times p} e^{-i p \times p'}	
 \right) \\
\end{array}
\end{equation}
Now the dependence on the cutoff scale $M$ is retained to illustrate IR/UV mixing explicitly: if we take
$\theta \to 0$ or the photon momentum $q \to 0$ before taking $M \to \infty$, the terms in this equation vanish
 ($M_{eff} \to M$, and both
$C$ and $\tilde{C}$ are proportional to $q^2 \theta^2$). But if
we take the ultraviolet limit  $M \to \infty$ first, then $M^2_{eff} \to 1/q\circ q$ (cancelling
the $q^2 \theta^2$-dependence in $C$ and $\tilde{C}$), and these
terms are now finite and nonzero as $\theta \to 0$ or $q \to 0$. Although this problem does not
necessarily hinder phenomenological analysis in a $\nc$ theory since in either order
 of limits one obtains finite ${\cal O}(\theta)$
corrections to the commutative theory, it is nonetheless a theoretical problem that needs further
attention.  

We note that not every situation requires renormalization with $\theta$-dependent
counterterms. Take for example
$K-\overline{K}$-mixing,
where the relevant loop integral is approximately\cite{Hinchliffe:2001im}
 \begin{equation}
\label{ncloop}
i \lambda^3 \rho \int{  d^4 k ~ \frac{ (\frac{k^2 m_q^4}{4 m_W^4} + k^2 - \frac{2 m_q^4}{m_W^2})
		\abs{k}\abs{p_1 \cdot \theta \cdot p_2} } {(k^2-m_q^2)^2 (k^2-m_W^2)^2 }}
\end{equation}
This is finite for all $\theta$ and external momenta.

\section{Phenomenology}
\label{sec:phen}
\subsection{New Frames of Reference for Particle Experiments}
 Since $\theta$ may have a preferred direction, experiments 
sensitive to $\ncg$ may therefore be measuring the components of
 $\overrightarrow{\theta}$, and it is necessary to take into account
the motion of the lab frame in this
 measurement\cite{Kamoshita:2002wq} .  Since $\nc$ effects are measured in powers of
$p^{\mu}   \theta_{\mu \nu} p'^{\nu}$, where $p,p'$ are some momenta involved
in the measurement, it is possible that odd powers of $\theta$ will partially average
to zero if the time scale of the measurement is long enough. Effects of first order in $\theta$
vanish at a symmetric $e^+ e^-$-collider, for example, if the measurement averages 
over the entire $4\pi$ solid angle of decay products. If the data is binned by angle
then it is possible to restore the sensitivity at ${\cal O}(\theta)$.
In addition to any other averaging process over short time scales, 
terrestrial
 experiments performed over several
days will only be sensitive to the projection of  $\overrightarrow{\theta}$
on the axis of the Earth's rotation. 
 Of course binning the data hourly or
at least by day/night, taking into account the time of year, can partially mitigate this effect.
This axis, as well as the motion of the solar system, galaxy, $\etc$, does not vary over time scales 
relevant to terrestrial experiments. In the subsequent review of the phenomenology, this consideration
of frame of reference should be held in mind.

\subsection{High-Energy Scattering Experiments}
\label{subsec:scatt}
One would expect effective operators involving $\theta$ (mass
dimension -2) to be
most relevant at high energies. Leptonic experiments may be possible
at a future collider with $\sqrt{s} = 500GeV$\cite{Aguilar-Saavedra:2001rg,
Abe:2001nr,Assmann:2000hg} and 
hadronic experiments will occur at  $\sqrt{s} \approx 10~TeV$ \cite{:1995mi}.
These will  constrain $\theta$. 

\subsubsection{Electron Scattering}

\begin{figure}[t]
\dofig{6.00in}{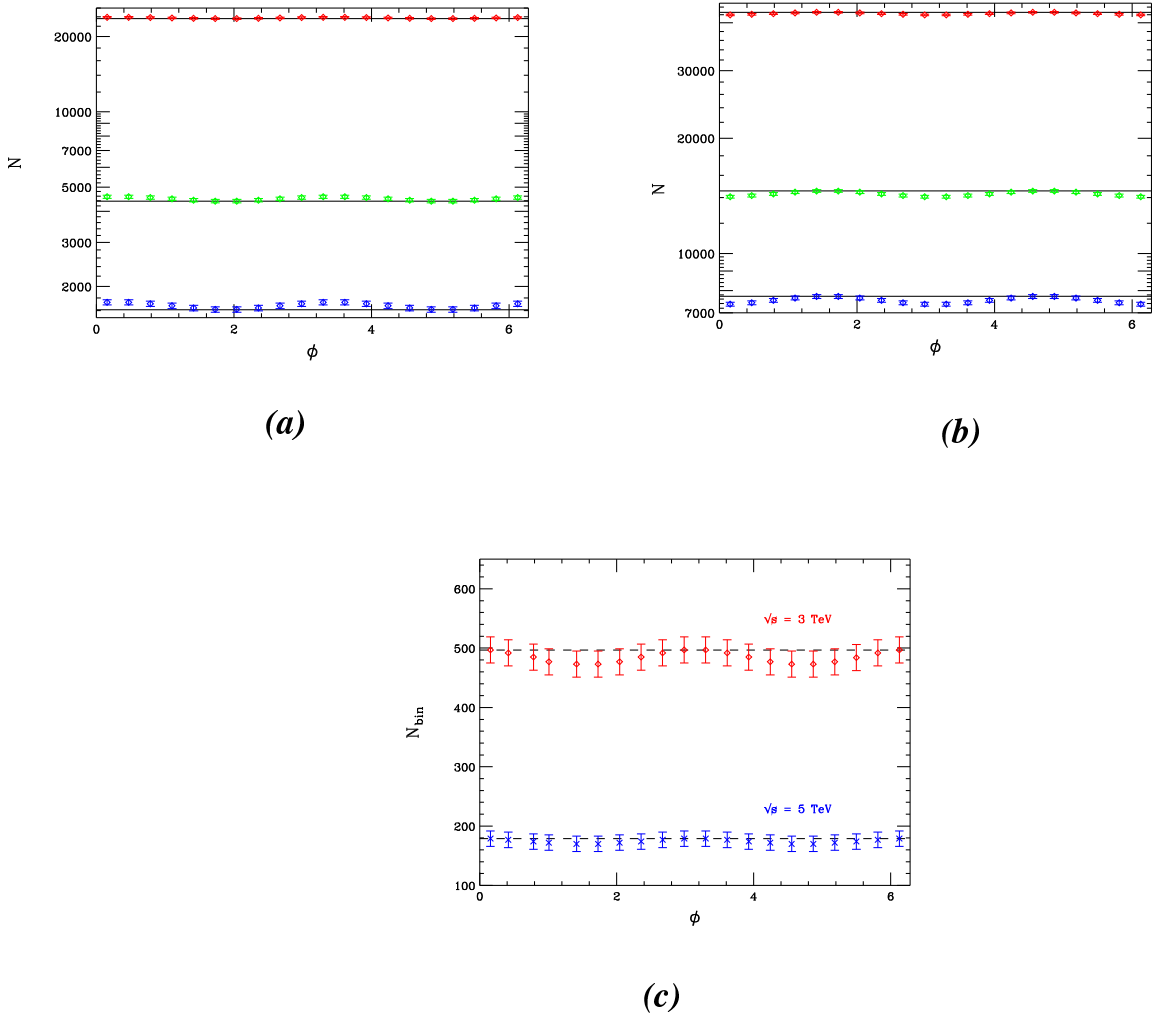}
\caption{Prediction of electron scattering differential cross sections in the azimuthal angle
$\phi$ in \cite{Hewett:2002zp} for (a) Bhabha scattering with
$\theta_{02}=(3~TeV)^{-2}$ and (b) M\"{o}ller scattering with $\theta_{12}=(3~TeV)^{-2}$
at a $\sqrt{s}=3~TeV$ linear collider with $1~ab^{-1}$ integrated
luminosity. Cuts on the polar angle $\abs{cos(\theta)}=0.9,0.7,0.5$ (top to bottom) are shown.
(c) Pair annihilation at $\sqrt{s}=3~TeV$ or $\sqrt{s}=5~TeV$, $\theta_{02}=1/s$.
 In all cases the solid lines are the SM prediction.
\label{escatter} }
\end{figure}

There are already a number of predictions of what one might observe in
a high energy $e^+ e^-$ scattering experiment 
\cite{Hewett:2002zp,Arfaei:2000kh,Hewett:2000zp,Mathews:2000we,Baek:2001ty,Mahajan:2001mf,Rizzo:2002yr,Godfrey:2001sc,Grosse:2001xz,Grosse:2001ei,Godfrey:2002yk}.
Such studies carry some degree of robustness since NCQED is the simplest
 $\nc$ extension of the Standard Model;
there is no need to define electroweak gauge interactions (see \ref{subsec:gauge} above)
as long as one only works with electrons, positrons, and photons (quarks aren't permissible
in the simplest scheme since they carry {\it fractional} charge, which wouldn't be a fundamental
representation of $U(1)$ in the same theory electrons existed in). The conclusion of the work is
that $e^+ e^-$ cross sections depend most sensitively on the 
projection of the particles' momenta on the plane perpendicular 
to $\overrightarrow{\theta}$. For example, if the beam is in the 1-direction,
then cross sections will not be sensitive to $\theta_{23}$ (and $\theta_{32}$).
 Sensitivity to the other
components of $\theta_{\mu\nu}$ depends on the process under consideration.
The authors in \cite{Hewett:2000zp,Hewett:2002zp,Rizzo:2002yr} find that differential cross sections for 
 M\"{o}ller scattering depend chiefly on $\theta_{13}$ and $\theta_{12}$ (the ``space-space''
components, due to t- and u- channel
interference) and show a statistically significant azimuthal dependence in a $500~GeV$ 
linear $e^+ e^-$ collider with $300~fb^{-1}$ integrated luminosity, assuming the
scale of noncommutivity is a few $TeV$.
 There is moreover a periodic 
behavior in the total cross section as the $\sqrt{s}$ of the machine changes.
Bhabha scattering and pair annihilation processes on the other hand test $\theta_{0i}$ (now the s-channel
plays a role in the interference), $\ie$ it is the ``space-time'' components of $\theta_{\mu\nu}$. 
Since it is likely that $\nc$ effects lie above 
$1~TeV$(see Sec \ref{sub:low} below),
we reproduce in Figure \ref{escatter} more recent predictions
 in \cite{Hewett:2002zp} for the azimuthal dependence of various cross sections in a 
linear  $e^+ e^-$ collider operating at $3~TeV$ or more with $1~ab^{-1}$ integrated luminosity,
 assuming $1/\sqrt{\theta} \approx \sqrt{s}$.

\begin{figure}[t]
\dofig{4.00in}{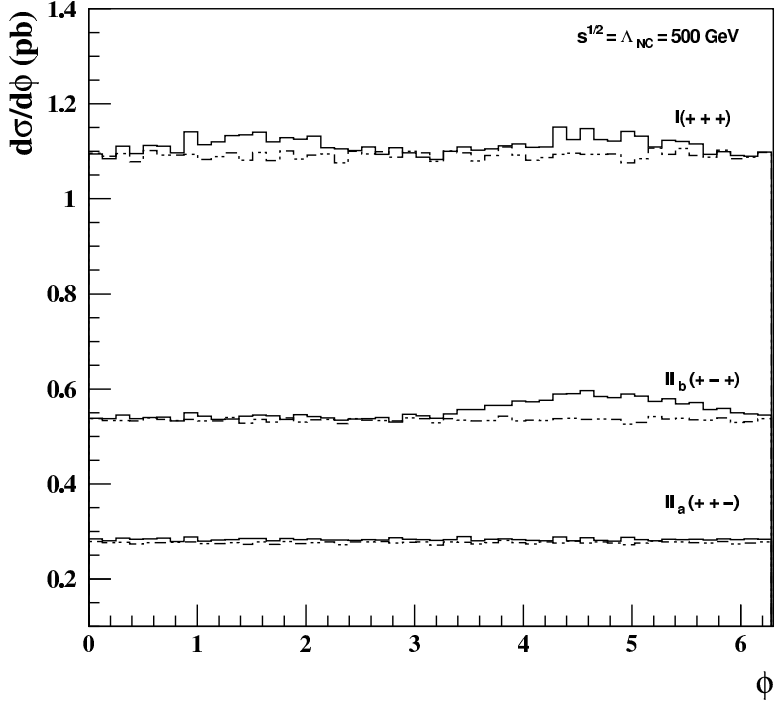}
\caption{$e-\gamma$ scattering at a $\sqrt{s}=500~GeV=1/\sqrt{\theta}$ linear collider
considered in \cite{Mathews:2000we} for space-space (I) and space-time (II) noncommutivity and
  polarization states of the (initial electron, photon, and final
electron). Here the dotted lines are the SM prediction, and $\Lambda_{NC} \equiv 1/\sqrt{\theta}$.
\label{egammafig} }
\end{figure}

\begin{figure}[t]
\dofig{4.00in}{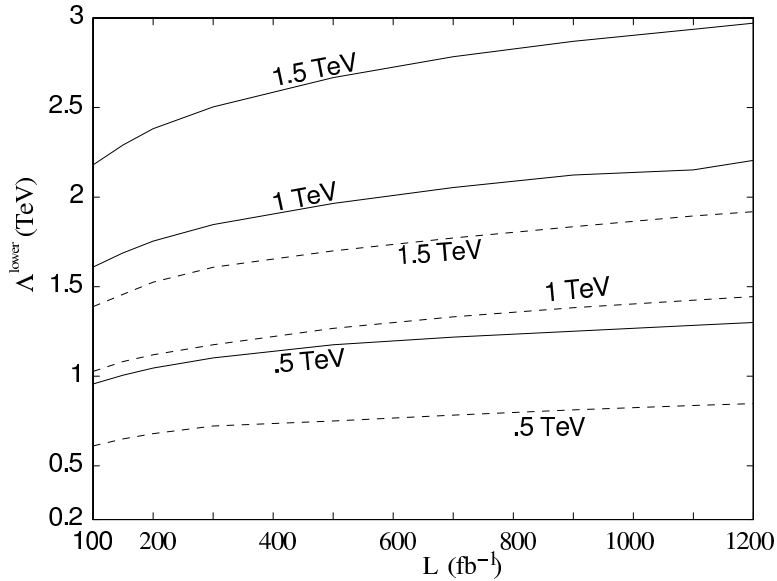}
\caption{Limits on $1/\sqrt{\theta}$ (95\% C.L.) from $\gamma-\gamma$ scattering at an $e^+-e^-$ linear collider
       considered in \cite{Baek:2001ty}, at various luminosities.
 Solid lines are for monochromatic photons, which are difficult to obtain in
practice. Dotted lines are for using backscattered photons. The $\sqrt{s}$ under consideration is listed
next to each line.
\label{gammafig} }
\end{figure}

 In a high energy $e^+ e^-$ linear collider, it is possible to backscatter laser beams off the
electron and positron beams, making $e \gamma$ or $\gamma \gamma$ 
collisions. The author of \cite{Mathews:2000we}, for example, finds $e \gamma$ scattering at a
$500~GeV$ linear collider with $500~fb^{-1}$ integrated luminosity can measure or exclude $1/\sqrt{\theta}< 1~TeV$
at 95\% C.L. In Figure \ref{egammafig} we show an azimuthal angle distribution computed from that work.
Similarly, $\gamma\gamma$ scattering can provide constraints of comparable strength, as considered in
\cite{Baek:2001ty}.  Figure \ref{gammafig} shows the result of the analysis in that paper. Higgs production as per $\gamma \gamma \to H^0 H^0$, strictly
forbidden in the SM, can yield up to a $100~fb$ cross-section at a
$1.5~TeV$ linear collider if the Higgs mass is less than
 $200~GeV$\cite{Aliev:2002rf}. 

\subsubsection{Hadronic Scattering}
 At present no work has been done in
hadronic $\nc$ phenomenology, although it
 is possible to measure the effects of a $\nc$ space-time in hadronic processes. From 
the Feynman rules in Figure \ref{feyn} we see that gluon couplings are generically
 $\theta$-dependent, so one would expect some observable effect in jet production rates and
angular distributions in a hadron
collider, for instance. If the theory were managable (see Sec. \ref{subsec:gauge} above 
with respect to $SU(N)$ groups)
 this would provide a non-trivial 
consistency check through the precise dependence of the gluon couplings on $\theta$\cite{Carlson:2001sw}. 
Uncertainties in calculating QCD rates in hadron colliders
would make the small effects of $\nc$ physics harder to detect than those in $e^+e^-$ colliders.
This would be compensated by the larger available energies.

\subsection{Low-Energy and Precision Experiments}
\label{sub:low}
Low-energy experiments (center of mass energies $E \leq 1~GeV$
tend to produce weaker limits on $\theta$ 
(the effect is quadratically suppressed as $E^2 \theta$) unless the sensitivity of 
the experiment is extraordinary. We begin with atomic and dipole moment constraints, as
these are among the most precisely measured quantities in physics.

\subsubsection{Atomic Transitions}
The authors in \cite{Chaichian:2000si,Chaichian:2002ew} apply $\nc$ quantum mechanics (NCQM) to the hydrogen atom 
to calculate the Lamb shift. In NCQM one replaces the position operator $x_i$ with
the $\theta$-deformed $x'_i = x_i + \frac{1}{2 \hbar} \theta_{ij}p_j$ and then applies the usual
rules of QM. In particular, the Coulomb potential is modified to $V(r) = -\frac{e^2}{r}
-e^2\frac{(r\times p)\cdot \overrightarrow{\theta}}{4 \hbar r^3} + {\cal O}(\theta^2)$.
They find the constraint on $\theta$ is: $1/\sqrt{\theta} \geq 10~TeV$.  The Stark and Zeeman effects 
add nothing to this constraint. The physics of positronium being very similiar
to hydrogen, in \cite{Haghighat:2001ia,Caravati:2002ax} the positronium 
$2^3 S_1 \to 2^3 P_2$ splitting is also considered; 
the strongest constraint is $\theta \leq 10^{-5}\lambda_e^2$, 
 or $1/\sqrt{\theta} \geq GeV$. Transitions in the Helium atom give a
stronger constraint: $1/\sqrt{\theta} \geq  30~GeV$  \cite{Haghighat:2002pt}.

\subsubsection{New CP-violation}
Beyond the standard source of CP violation from the mismatch between the mass and weak
quark (and possibly lepton) eigenstates, in the NCSM there is an additional source of $\cpviol$:
the parameter $\theta$ itself is the $\cpviolng$ object. To see this, consider (as in \cite{Sheikh-Jabbari:1999iw})
 the 
transformation of the action (\ref{ncqed}) under the discrete symmetries $P,C$. Under the parity
transformation,
\begin{equation}
\label{parity}
\begin{array}{lll}
x_i & \to & -x_i \\
A_0 & \to & A_0 \\
A_i & \to & -A_i \\
\psi(x) & \to & \gamma^0 \psi(x) \\
\theta_{\mu\nu} & \to & \theta_{\mu\nu} \\
\end{array}
\end{equation}
leaves the action invariant. However under charge conjugation
\begin{equation}
\label{charge}
\begin{array}{lll}
\partial_\mu & \to & \partial_\mu \\
A_\mu & \to & -A_\mu \\
\theta_{\mu\nu} & \to & -\theta_{\mu\nu} \\
\end{array}
\end{equation}
is required to keep the field strength tensor in (\ref{Fterm}) unchanged.
Therefore, under $C$ and $P$ combined $\theta \to -\theta$.
More detailed work reveals that $\theta$ is 
in fact proportional to the size of an effective particle dipole moment
\cite{Ardalan:1998ce}. Therefore
$\ncg$ can actually {\it explain} the origin of $\cpviol$. 

\subsubsection{Electron Electric Dipole Moment}
Since the SM predictions of the $\cpviolng$ electric dipole moments ($\edm$s) 
are extremely small, we might expect that new 
sources of $\cpviolng$ physics from $\ncg$ would be observable. The 
noncommutative geometry  provides
in addition a simple explanation for this type of 
$\cpviol$: the directional sense of the $\edm$ $\overrightarrow{D}$ derives from
the different amounts of noncommutivity in different directions 
($\ie$ $D_i \propto \epsilon_{ijk}\theta^{jk}$) and the 
size of the $\edm$, classically proportional to the spatial extent of a charge
distribution, is likewise in $\ncg$ proportional to  $\sqrt{\theta}$, the
inherent ``uncertainty'' of space.
The effects of $\ncg$ will be proportional to the typical momentum
involved, which for an electron $\edm$ observation is $\sim keV$. A detailed
analysis of the size of the $\edm$ appears in
 \cite{Riad:2000vy, Iltan:2003vi}. The magnetic 
dipole moment, incidentally, receives a very small and (to leading order
in $\theta$) spin-independent 
contribution from $\ncg$ which makes it extremely difficult to observe.
A simple estimate of the expected electron electric dipole moment\cite{Hinchliffe:2001im} yields
a fairly strong bound: $1/\sqrt{\theta} \geq 100~TeV$.
However, we cannot exclude the
possibility that the actual $\edm$ is much smaller than the above naive
estimate, a situation which can arise in supersymmetric models \cite{Barger:2001nu,
Brhlik:1998zn}. 

\subsubsection{CP Violation in the Electroweak Sector}

At the field theory level, it is the momentum-dependent phase factor appearing in the 
\newline
 $\nc$
theory which gives $\cpviol$. For example, the NCSM W-quark-quark $SU(2)$ vertex
in the flavor basis is
\begin{equation}
\label{wqq}
{\cal L}_{Wqq} =  \overline{u(p)} \gamma^\mu (1-\gamma_5) e^{ip\cdot \theta \cdot p'}~d(p')~W_\mu  
\end{equation}
Once we perform rotations on the quark fields to diagonalize the Yukawa interactions, 
$\ie$ $u_L \to U u_L$ and $d_L \to V d_L$, the above becomes
\begin{equation}
\label{wqqmass}
{\cal L}_{Wqq} =  \overline{u(p)} \gamma^\mu (1-\gamma_5) 
e^{ip\cdot \theta \cdot p'} U^\dagger V ~d(p')~W_\mu  
\end{equation}
Even if $U^\dagger V$ is purely real, there will be some nonzero phases $e^{ip\cdot \theta \cdot p'}$ 
in the Lagrangian whose magnitudes increase as the momentum flow in the process increases. Of course,
the above phase factor has no effect at tree-level (suitably redefining all the fields) but will
affect results at 1-loop and beyond. The overall effect is to introduce momentum-dependent 
phases into the CKM matrix. The authors in \cite{Hinchliffe:2001im} consider such a matrix:
\begin{equation}
\label{ncCKM}
\overline{V}(p,p') \equiv  
	  \left( \begin{array}{ccc} 
	1- \lambda^2/2 + ix_{ud} & \lambda  + ix_{us}  &
         A \lambda^3 (\rho-i\eta) + ix_{ub} \\
	-\lambda +  ix_{cd}& 1- \lambda^2/2 + ix_{cs} &
         A \lambda^2 +  ix_{cb} \\
	 A \lambda^3 (1-\rho-i\eta)+  ix_{td}  & -A \lambda^2 +  ix_{ts} &
          1 +  ix_{tb}\\ \end{array} \right)
\end{equation}
where $x_{ab} \equiv {p_a}^\mu \theta_{\mu \nu} {p'_b}^\nu$ for quarks $a,b$. 
This matrix is an approximation of the exact ncSM in the perturbative limit
where we expand
$ e^{ip\cdot \theta \cdot p'} \approx 1 + ip\cdot \theta \cdot p'$.
In the limit $\theta \to 0$, the $x_{ab}$ all go to zero and $\overline{V}$
becomes the CKM matrix $V$ in the Wolfenstein parameterization\cite{Wolfenstein:1983yz} in terms of the
small number $\lambda \approx 0.22$. 

\begin{figure}[t]
\dofig{4.00in}{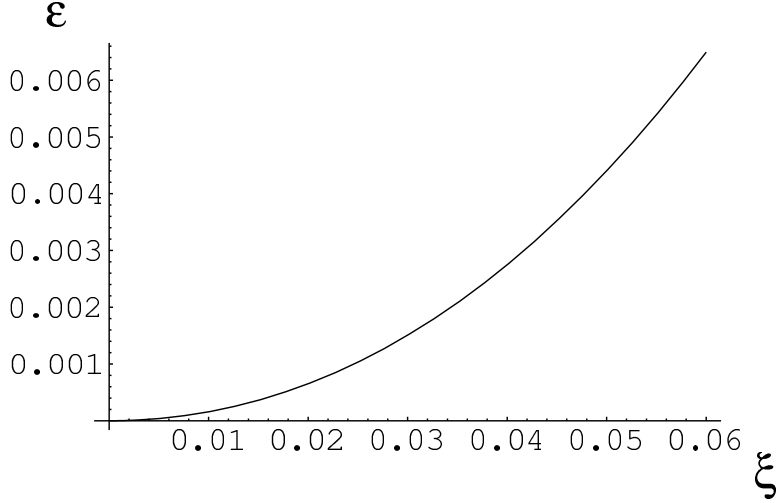}
\caption{CP-violating effects of $\ncg$: if the Wolfenstein parameter $\eta=0$, $\ncg$ alone could
account for $\epsilon_K \approx 2.28\cdot 10^{-3}$ if $\xi \approx 0.04$ 
($\xi \equiv M_W \sqrt{\theta}$, so this corresponds to $1/\sqrt{\theta} \approx 2~TeV$).   
\label{cpfig} }
\end{figure}

The authors in \cite{Hinchliffe:2001im} calculate whether the 
$\theta$-dependent phases are observable in $K-$ or $B-$ physics: the conclusion is that
if $1/\sqrt{\theta} \approx 1~TeV$ then $K-$physics observables such as $\epsilon_K$ and 
$\epsilon'/\epsilon$ are sensitive to $\ncg$ (see Figure \ref{cpfig}), however $B-$physics observables, such
as $sin2\beta$, are not.
However in $D-$physics, the asymmetry ${\cal A}_s$ between the rates
 $D^-_s \to K^- K_s$ and
 $D^+_s \to K^+ K_s$ could provide a measurable signal. The authors in 
\cite{Chang:2002ix} calculate
\begin{equation}
{\cal A}_s \equiv \frac{\abs{A(D^-_s \to K^- K_s)}^2 - \abs{A(D^+_s \to K^+ K_s)}^2   }
{\abs{A(D^-_s \to K^- K_s)}^2 + \abs{A(D^+_s \to K^+ K_s)}^2}
\approx 2 \Re {\epsilon_K} - 2 {\cal J} R_s sin \delta_s
\end{equation}
where ${\cal J} \approx A^2 \lambda^6 \eta - \lambda^2 p^2 \theta$ ($p$ being the
typical momentum flow in the process), $R_s \approx -1.2$,
and $\delta_s$ is a phase from the strong dynamics. 
With $p \approx M_{D^+} \approx 1.97~GeV$\cite{Groom:2000in}, a measurement of this
asymmetry would be sensitive to $\nc$ effects for $1/\sqrt{\theta} \le 1~TeV$.
 
Other work in $\cpviol$ includes the enhancement of the decay $\pi \to 3\gamma$ relative to the SM
by many orders of magnitude \cite{Grosse:2001ei}. Unfortunately, the branching ratio for this decay,
assuming $1/\sqrt{\theta} \approx 1~TeV$, is ${\cal O}(10^{-20})$, whereas current experimental bounds
are at ${\cal O}(10^{-8})$.

\subsubsection{$(g-2)\mu$}

\begin{figure}[t]
\dofig{2.00in}{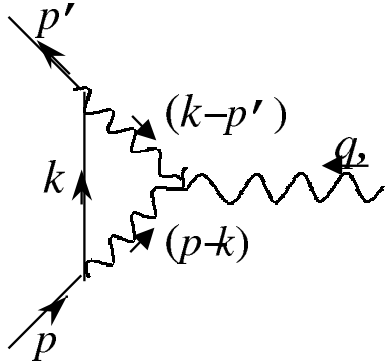}
\caption{A novel contribution to the muon's magnetic moment from $\ncg$ considered in \cite{Wang:2001ig}.
\label{muonfig1} }
\end{figure}

Although the leading order contribution to a fermionic magnetic moment 
doesn't couple to the fermion's spin (and therefore not leading
to precession in an external magnetic field,
making it hard to observe), at
 second order in $\theta$ this is no longer the case. There is a novel contribution
to the magnetic moment of the muon coming from the diagram shown in Figure \ref{muonfig1}.
	In an experiment such as the one at BNL \cite{Brown:2001mg}, where muons
circulate in a ring at a very high momentum ($\gamma \approx 30$),
 the second-order contribution to the anomalous magnetic moment of the muon has been
calculated\cite{Wang:2001ig} in NCQED to be
roughly of order 
$\delta a_\mu \approx \alpha {m_{\mu}}^2 \theta^2 {E_\mu}^2/96\pi \gamma^4$ which,
considering that the discrepancy between experiment and theory may be as large
as $\delta a_\mu \leq 10^{-9}$ \cite{Czarnecki:1998nd}, 
suggests a lower bound $1/\sqrt{\theta} \geq m_\mu$.

\begin{figure}[t]
\dofig{5.00in}{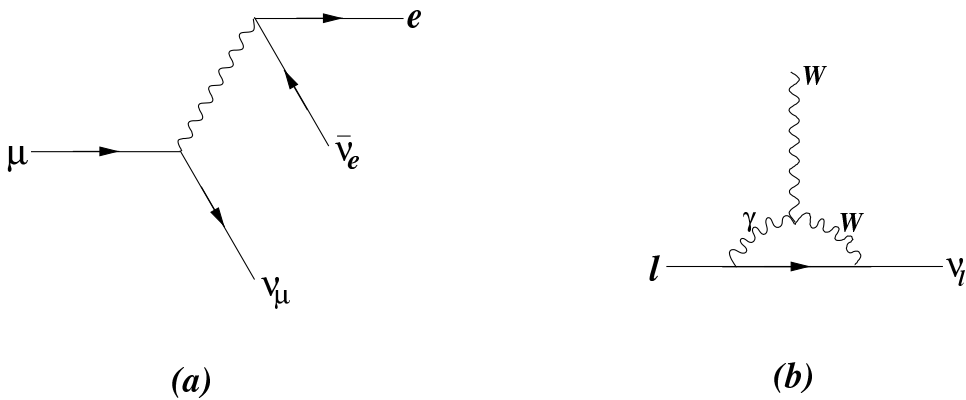}
\caption{(a)Tree-level muon decay occurs through the exchange of a $W$-boson.  When the
electron is emitted with an energy close to its kinematical upper bound of $m_\mu/2$ in the
center-of-mass frame,
the neutrinos are ejected from the decay vertex in the opposite direction and the spin of
the electron becomes highly correlated with the spin of the muon. (b) The 
$W-l-\nu$ vertex receives a $\theta$-dependent one-loop correction which can upset the
aforementioned spin correlation \cite{Kersting:2001zz}.
\label{muonfig2} }
\end{figure}

There is a much stronger bound in the literature which is derived from the
 ${\cal O}(\theta)$ correction to muon decay. The BNL experiment measures
$a_\mu$ by comparing the cyclotron frequency to the
observed precesion rate of the muon's spin; the difference between these frequencies
is directly proportional to $a_\mu$. In this
experiment the rate of precession of the muon's spin
 is measured indirectly from the decay of 
the muons into electrons.
 Electrons emerge from
the decay vertex with a characteristic angular distribution which
in the Standard Model (SM) has the following form in the rest frame of the muon:
\begin{equation}
\label{asymm}
dP(y,\phi) = n(y) (1 + A(y) cos(\phi))dy d(cos(\phi))
\end{equation}
where $\phi$ is the angle between the momentum of the electron
$e$ and the spin of the muon, $y = 2 p_e/m_\mu$ measures the fraction of the maximum 
available energy which the electron carries, and $n(y),A(y)$ are 
particular functions which peak at $y=1$. The detectors (positioned 
along the perimeter of the ring) accept the passage of only the highest
energy electrons in order to maximize the angular asymmetry 
in (\ref{asymm}). In this way, the electron count rate is
modulated at the frequency $a_\mu e B/(2\pi m c)$, the difference between the cyclotron and precession
frequencies of the muons. 
However each of the $W-e-\nu$ vertices in the decay diagram receives a one-loop correction
which is $\theta-dependent$, assuming the validity of a NCSM (see Sec \ref{sub:work} above).
The appearance of the antisymmetric object $\theta_{\mu \nu}$
in the decay amplitude leads to combinations of the muon and electron spins 
and momenta which alter the modulation frequency of the decay rate (\ref{asymm}).
Specifically, one anticipates factors of 
($\overrightarrow{p_e} \cdot \overrightarrow{s}_\mu)
(\overrightarrow{p_e} \cdot \theta \cdot \overrightarrow{s}_\mu)$
which for electron momenta close to their kinetmatical limit ($\ie$ $y=1$)
behaves like $cos(\phi)sin(\phi)$ and upsets the simple angular dependence in (\ref{asymm}).
In \cite{Kersting:2001zz} the $\theta$-dependence of the decay is calculated and based on
the degree to which this changes the apparent precession of muon spin the
bound $1/\sqrt{\theta} \geq 1~TeV$ is obtained. These bounds are subject to improvement in
light of the ongoing analysis of the BNL data \cite{Chattopadhyay:2002zq}.

\subsubsection{Contributions from Two Loops}
\label{subsub:nuc}

The authors in \cite{ Anisimov:2001zc} derive a two loop contribution to the effective NCQED action
of the form
\begin{equation}
{\cal L}_{eff} = \frac{3}{4}m \Lambda^2 \left(\frac{e^2}{16 \pi^2}\right)^2 \theta^{\mu\nu} {\overline \psi}
	\sigma_{\mu\nu} \psi
\end{equation}
where $\Lambda$ is an effective cutoff. They conclude that if $\Lambda \sim 1~TeV$ then experiments sensitive
to background magnetic fields could constrain $\theta < (10^{12-13} GeV)^{-2}$. Such experiments may
include the variation
of spin precessions with time \cite{  Carlson:2001sw,Lane:2002ku}.
 Experiments typically look
for sidereal variations of electronic or nuclear spin, for example in
Cs or Hg, the latter's nuclear spin giving
the stronger constraint. If $\theta$ is constant in space over distances comparable
to those the Earth moves through on a timescale of months, then its direction
behaves as a background magnetic field, possibly giving rise to effective operators like
$\theta^{\mu\nu}\overline{q}\sigma_{\mu\nu}q$ which would cause noticable variations in
the nuclear spin precession frequencies (observed on scales of months).
 To the extent that such variations are on the order of 
microhertz or hundreds of nanohertz\cite{Hunter:1998ej,Bear:2000cd}, bounds on the 
 projection of $\overrightarrow{\theta}$ on the axis of the 
Earth's rotation (see Section \ref{subsubsec:lorentz} above)such as 
$1/\sqrt{\theta} \geq 10^{14}~GeV,10^{17}~GeV$ are obtained\cite{Mocioiu:2001nz}. Such bounds
are on a weaker footing than the ones from NCQED as the complete formulation 
of NCQCD has not been thoroughly developed\cite{Carlson:2001sw}.

\subsubsection{Other Precision Tests}
Among the other ideas for testing noncommutivity are directly measuring the dispersion 
relation for photons, which is altered in a $\nc$ space-time, in a Michelson-Morley-type
interferometry experiment \cite{Guralnik:2001ax}. Those authors find that in principle one
could be sensitive to $1/\sqrt{\theta} \leq 10~TeV$ in an interference experiment with visible light
in which the sum of the legs of the apparatus extended for one parsec in a background
magnetic field of 1 Tesla; they note this 
distance, however, is impractically large to coherently maintain such a strong background field.
Moreover there is the question of the uniformity of $\theta$ over these distances.
A similar bound is obtained from considering an Aharonov-Bohm effect with high
energy electrons\cite{Falomir:2002ih}.
Reference \cite{Iltan:2002fa} discusses $\nc$ contributions to $b \to s \gamma$ 
inclusive decays.
In another paper \cite{Deshpande:2001mu} the effects of $\ncg$ on the triple neutral
 gauge boson couplings is considered. They use an $SU(5)$ unification scheme to
motivate a particular choice of gauge transformation parameters left undetermined
in the methods that \cite{Calmet:2001na} employ to construct consistent
$\nc$ non-Abelian gauge interactions in the NCSM. 
At first order in $\theta$, they find analytical contributions to 
couplings ($e.g.$ $\gamma\gamma\gamma$, $\gamma\gamma Z$, $\gamma g g$ ...)
which are strictly forbidden in the SM.  The work in \cite{Behr:2002wx} follows up on
this and derives the partial width $Z \to \gamma\gamma$ in the NCSM; this decay
has a branching ratio of $10^{-10}$ in the SM (current experimental bounds are  $B.R.<10^{-4}$\cite{Groom:2000in})
 so its observation could be a clear positive
indicator for $\ncg$. High energy processes in stars may also provide some useful constraints: in \cite{Schupp:2002up} a limit of $1/\sqrt{\theta} \geq 80~GeV$ is placed on the noncommutivity scale based on the criterion that any additional energy loss from the star due to noncommutivity not exceed the standard amount due to neutrinos. Some theoretical
 work\cite{Kokado:2002sz} on the Hall effect indicates 
that one can test noncommutivity here, though as yet there is no direct
comparison with exeriment.  Finally, \cite{Castorina:2002vs}
 presents a calculation of 
the effects of noncommutivity on synchrotron radiation and, although no 
numerical bounds are presented, asserts this could be a good way to 
search for $\nc$ effects.

\subsection{Cosmology and Extra Dimensions}
\label{subsec:cosmosdim}

Finally, we will say a word about developments in $\nc$ cosmology and extra dimensions.
Noncommutative effects in cosmology range from birefringence of light, a variable
speed of light, an alterted cosmic microwave background, highly energetic photons and protons above the 
GZK\cite{Greisen:1966jv,Zatsepin:1966jv} cutoff (that is, the threshold on cosmic proton energies due to the
reaction $p+\gamma \to p+ \pi$ with the cosmic microwave backround radiation (CMBR)), to the overall quantum 
structure of the cosmos \cite{Garcia-Compean:2001wy,Cai:2001az,Alexander:2001ck,Tamaki:2001ck,Lizzi:2002ib}.
In \cite{Tamaki:2001ck}, for example, $\ncg$ is a possible explanation of the detection of highly energetic
photons ($E>20~TeV$ from sources $>100~Mpc$ distant \cite{Aharonian:2002cv}). It is a mystery how such energetic photons
can reach detectors on Earth since the interaction with the infrared background via $\gamma \gamma \to e^+ e^-$
should impose a threshold energy of $\approx 1~TeV$ on photons travelling over galactic distances. In considering a 
$\theta$-deformed spacetime, the authors of \cite{Tamaki:2001ck} derive modified relativistic kinematics introducing a 
breaking of Lorentz invariance. Their results indicate that the threshold can be pushed to
much higher values if $1/\sqrt{\theta} <= 1~TeV$. The same analysis can account for the recent observation of highly energetic 
cosmic rays far above the GZK cutoff, a limit of about $7 \cdot 10^{19} eV$ on the energy protons can have 
due to the interaction with the CMBR.  A $\nc$ scale as high as $10^8 TeV$
can push the threshold above the observations.
   
The reason why we expect $\ncg$ will have
a large effect on a cosmic scale is two-fold. First, large distances are available between
source and observer over
which even a small change in a particle's dispersion relation can accumulate to a
produce a large effect. Second, the evolution of the cosmos went through an epoch
when temperatures were far above the $\nc$ scale $1/\sqrt{\theta}$: $\nc$ effects were non-negligable,
perhaps even dominant, and may have left a lasting trace on the global structure of the cosmos after
inflation. These are very interesting ideas to pursue, but also very difficult, as they involve
nonperturbative effects of $\ncg$. 

\begin{figure}[t]
\dofig{5.00in}{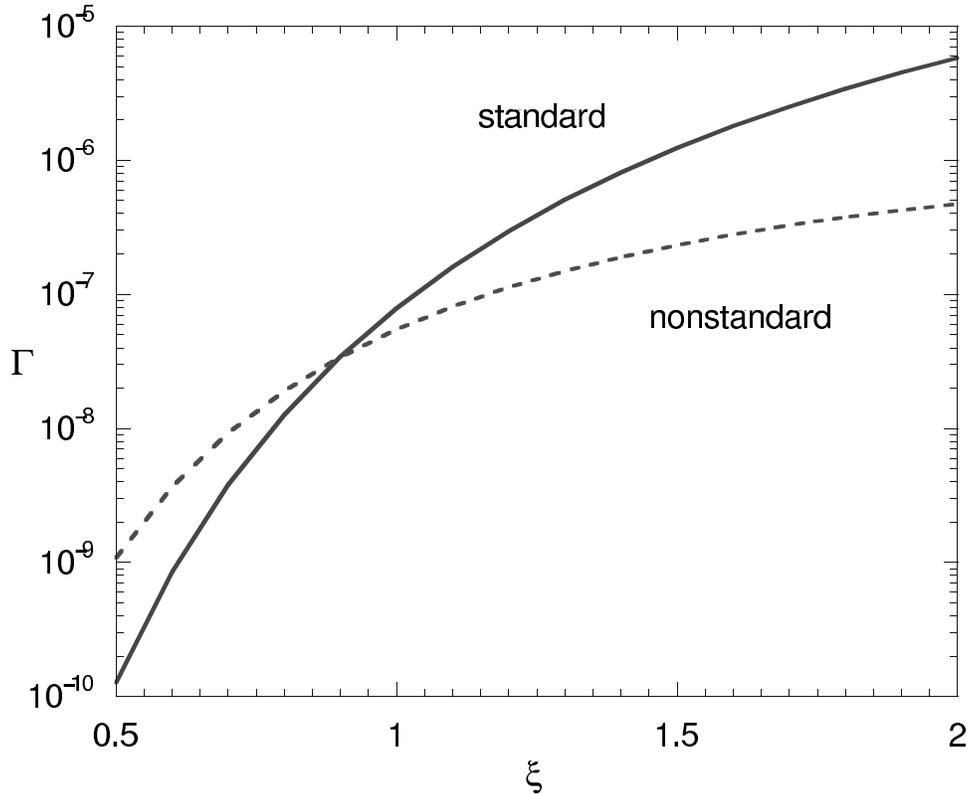}
\caption{The decay of a KK photon to a lighter KK photon plus $f\overline{f}$ pair 
in the standard (non-$\nc$) $T^2/Z_2$ extra dimensions scenario can receive corrections of
 over 100\% from $\nc$ effects in the non-standard $\nc$ extra dimensional scenario according
to the work in \cite{Carlson:2001bk}. Partial width is in units of the compactification scale $1/R$, and the
 parameter $\xi$ denotes the ratio of the compactification radii of the torus. 
\label{xtrafig1} }
\end{figure}

Noncommutative extra dimensions may be a more attractive alternative to $\nc$ 4-space. The
authors in \cite{Carlson:2001bk} consider six-dimensional NCQED where the noncommutivity is confined
to two bulk dimensions compactified on a toroidal orbifold $T^2/Z_2$. Such a scenario 
explicitly evades the tight four-dimensional Lorentz-violating bounds mentioned above in connection with 
nuclear spins, because ordinary matter doesn't exist in the bulk.
They find that for heavier Kaluza-Klein (KK) modes (several $TeV$ or more) the decay
of a KK photon
to a fermion-antifermion pair,
$\gamma^{n_a} \to \gamma^{n_b} f \overline{f}$, may be observable at a Very Large Hadron
Collider(VLHC) \cite{vlhc}. Decay rates may be up to 100\% larger for noncommuting extra dimensions
than for ordinary commuting ones (see Figure \ref{xtrafig1}). We quote their claim that a $2~TeV$ initial state produced at
a $\sqrt{s}=200~TeV$ VLHC with $100~fb^{-1}$ integrated luminosity would enhance the production
rate to allow 
the discrimination of commuting versus noncommuting extra dimensions at a level of  $6.8~\sigma$. 
On the other hand, for lighter KK modes, a pair production process such as 
$f \overline{f} \to \gamma^{n_a} \gamma^{n_b}$ is the more favorable one to observe at a
collider such as the LHC.

\section{Challenges and Conclusions}
\label{sec:chall}

While the NCSM has some
technical difficulties, particularly with regards to the IR/UV problem
and in the gauge sector, it is possible to estimate the effects of such
extensions in many cases as we have shown above. As discussed
 in Section \ref{subsec:qft},
versions of $\nc$ field theory involving supersymmetry are  easier to  
develop theoretically.

Apart from the issue
of the full theory there are other areas where
work on $\nc$ phenomenology is incomplete.  In particular, there is
as yet no compelling reason why $\theta_{\mu \nu}$ should be constant; a theory
with a
space-time dependent $\theta(x)_{\mu \nu}$ is feasible and is eventually
necessary
to investigate. In particular, if  $\theta_{\mu \nu}$ is space-dependent then
there is
no single direction it ``points'' in, and many of the stringent limits on
$\theta_{\mu \nu}$
which rely on this ``direction'' need reconsideration.               

Constraints on the NCSM arise from two distint classes of experiments:
very high precision measurments at relatively low energy that can see
indirect effects and
experiments whose energy scales approach that of the $\nc$ physics
that  can see the direct effects. The latter involves experimentation
at high energy colliders. In an $e^+e^-$ collider effects show up as
distortions in, for example the Bhabha, scattering rates. These
experiments are expected to be sensitive to $\sqrt{\theta}\sim
1/\sqrt{s}$.
  Moreover there is as yet no phenomenological
work on a $\ncg$ at the LHC or other hadron colliders. Indirect
constraints from measurements of magnetic moments can be very
stringent but their interpetation is model dependent.     

Finally, all the above ideas should be investigated more fully in the contex of
 $\nc$ cosmology and extra dimensions to the extent where
there are definite predictions. The work so far in these areas is encouraging.

\section*{Acknowledgements}
 This work was supported by the Director, Office of Science, Office
of Basic Energy Services, of the U.S. Department of Energy under
Contract DE-AC03-76SF0098, and by the Department of Physics at Tsinghua University.

\bibliography{all.bib}
\bibliographystyle{unsrt}

\end{document}